# Multi-plane, Multi-band image projection via Broadband Diffractive Optics

**MONJURUL MEEM,**[1] **APRATIM MAJUMDER,**[2] **AND RAJESH MENON**[2,3,*]

[1]*Department of Electrical and Computer Engineering, University of Utah, Salt Lake City UT 84112, USA.*
[2]*Oblate Optics, Inc., San Diego CA 92130, USA.*
*\*rmenon@eng.utah.edu*

**Abstract:** We demonstrate visible and near-IR image projection via non-absorbing, multi-level Broadband Diffractive-Optical Elements (BDOEs) in 1 or more planes. By appropriate design of the BDOE topography, we experimentally demonstrate: (1) different images in different spectral bands; (2) different images in different image planes; (3) image magnification by changing the distance between the illumination source and the BDOE; (4) completely flat BDOE via an index-contrast top-coating; and (5) reflective BDOEs. All of these are accomplished with broadband illumination. Furthermore, the BDOEs are highly efficient, versatile and can be inexpensively mass manufactured using imprint-based replication techniques.

## 1. Introduction

Broadband image projection can be achieved with absorbing color filters as in conventional image projectors. However, these are very inefficient. The same may be achieved with non-absorbing patterned dielectrics, which could be programmable (eg. spatial-light modulators) [1] or fixed (generally referred to as holograms or computer-generated holograms). Traditionally, these devices are sensitive to the wavelength of illumination, achieving high efficiencies only in narrow bandwidths (requiring lasers for their effective operation). Recently, there have been several attempts to improve the broadband performance of such devices either by using metamaterials [2,3] or via multi-level diffractive structures [4,5]. Metamaterials-based holograms require deep sub-wavelength minimum features and relatively large aspect ratios, which render them very challenging for practical applications and also are very inefficient, [6] especially when the mechanism is based upon plasmonics.

We previously demonstrated the design, fabrication and characterization of multi-level diffractive holograms that can project full-color images when illuminated by collimated white light [4,5]. The constituent element of such a broadband diffractive-optical element (BDOE) is typically a square pixel, whose minimum width is determined by the fabrication technology, and whose height is determined via a nonlinear optimization procedure. The nonlinear optimization is based on the modified direct-binary-search algorithm that we have described previously in detail [4,5]. The goal of optimization is to maximize the diffraction efficiency averaged over all the wavelengths of interest. In order to ensure manufacturability, we enforce additional constraints on the number of height levels. By exploiting the intrinsic chromaticity of diffraction as well as material dispersion, here we show that it is possible to achieve efficient image projection with temporally incoherent and spatially partially-coherent illumination. The specific advancements that are reported here are the projection of: a full-color image by illumination with the white-LED flashlight of a mobile phone, of two distinct images in the visible and in the invisible (near-infrared) bands, and of multiple distinct images in multiple planes. Intriguingly, we show the equivalence of our BDOE with a lens, whose point-spread function is structured as an image. This equivalence enables magnification of the projected images without any additional optics. We further showcase the versatility of BDOEs by demonstrating a completely flat device based on an index-contrast top-coating, and also a reflective BDOE. All the devices reported here have minimum feature width of 20μm (except one in Fig. 5, which has 10μm), maximum feature height of 2.6μm, and 100 levels. All other parameters are listed in Table S1 in ref [7]. However, it is important to note that BDOEs can be designed for any feature size as small as $\lambda/2$ (beyond which all free-space modes are evanescent). In contrast to diffraction gratings and far-field diffractive optical elements (including conventional holograms), our BDOEs are designed in the Fresnel regime. Therefore, the conventional diffraction efficiency (which is typically defined as a function of spatial frequency) is not meaningful. Instead, here we use the imaging efficiency (see ref [7] for details), which is a measure of the fidelity of the projected image, and is computed in real space [4]. In addition, we also report the measured transmission/reflection efficiency as a function of wavelength (> 96%), and confirm that our BDOEs also function as anti-reflection coatings for certain wavelengths.

## 2. Experimental Results for full-color, multi-band and multi-plane image projection

Each BDOE is comprised of square pixels, whose heights are determined by a nonlinear optimization method to maximize imaging efficiency [4]. The BDOEs are fabricated in a transparent photopolymer film (positive-tone photoresist S1813 from Microchem, with a maximum thickness of 2.6μm) placed atop a 2" soda lime wafer (thickness ~0.5mm). Direct laser write grayscale-optical lithography was used for patterning the BDOEs [4]. For characterization, the BDOEs were illuminated with an appropriate source and the projected images were recorded. For further details about the fabrication and experimental characterization of the BDOEs, please refer to ref [7].

Fig. 1 shows a BDOE that is designed to project a saturated-color image, when illuminated by the white LED flashlight of a mobile phone. The image was projected onto a white screen and photographed, and compared to the simulated image. Rich saturated colors spanning the entire visible range (blue, green, red) are clearly visible. As others have reported previously, [8] one can increase spatial coherence to improve image quality. Increasing the source-BDOE distance can increase the spatial coherence. In Fig. 1, we placed the phone (emission area diameter about 3mm) about 0.9m away from the BDOE to obtain collimated white light. At lower spatial coherence (closer source-BDOE distances), the same image is formed but with reduced sharpness as expected [8]. This device was also tested under direct sunlight and the same colorful image is observed (see Fig. S8 in ref [7]). The pixel heights for all the devices were measured using a stylus profilometer and these measurements confirm that the standard deviation of the error is less than 65nm (data included in section 3 in ref [7]). Our experiments confirm that one can use the flashlight of any mobile phone to project colorful images, which have not been demonstrated by meta-holograms or conventional holograms. We also note that all our devices are on-axis and exhibits negligible zero-order stray light in the projected images.

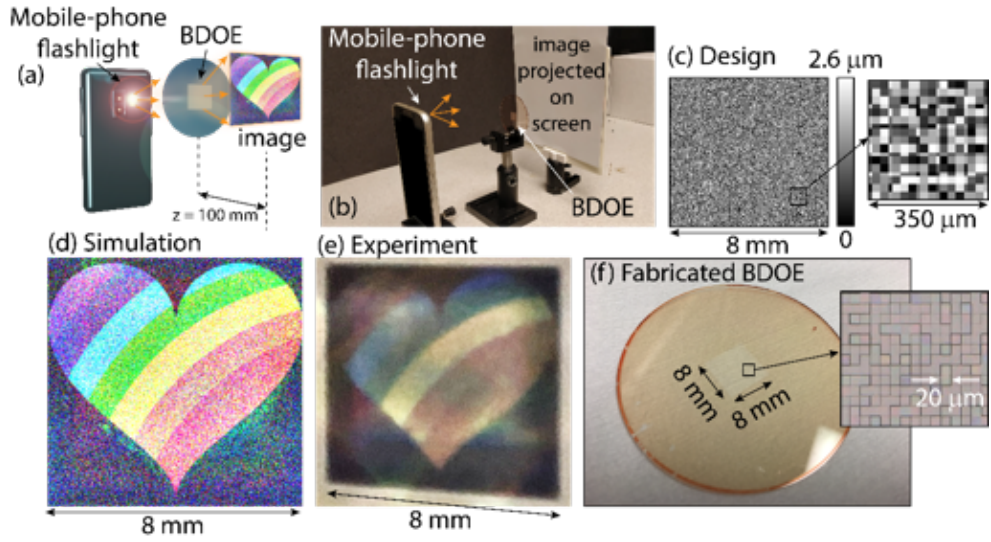

Fig. 1. BDOE illuminated by flashlight of a mobile phone. (a) Schematic and (b) photograph of the experimental setup. (c) Designed height map of the BDOE, showing (inset) magnified view of a small portion. (d) Simulated and (e) experimental full-color images when the BDOE is illuminated by the flashlight of a mobile phone. (f) Photograph showing a fabricated BDOE and (inset) optical micrograph of a small portion of the fabricated BDOE.

We can extend the technology to multiple spectral bands as illustrated in Fig. 2, where one BDOE produces either a full-color image or an invisible (Near-IR) image, when illuminated by a broadband white light (supercontinuum source with bandwidth of 400nm to 700nm, shown in Fig. 2b) or a near-infrared source (850nm laser, shown in Fig. 2c), respectively. We should mention that, this is the same BDOE presented in Fig. 1. It is noteworthy that there is very little cross-talk between the two spectral bands. We analyzed the crosstalk by computing the Structural Similarity (SSIM) index maps for the two planes and the global SSIM value, which for this case, was 0.027 and 0.085 in simulation and experiment, respectively. We have measured the imaging efficiency as well. The average imaging efficiency was measured to be 54%, which is in close agreement with the simulated efficiency of 64%. Further details are presented in ref [7]. This was predicted via simulations before, [5] but this is the first experimental demonstration. As a result, a single device enables both covert and overt security for anti-counterfeiting. In fact, under direct sunlight one can readily observe the rainbow heart and by simply placing a visible-cut filter also observe the IR image of the lion (see Visualization 1).

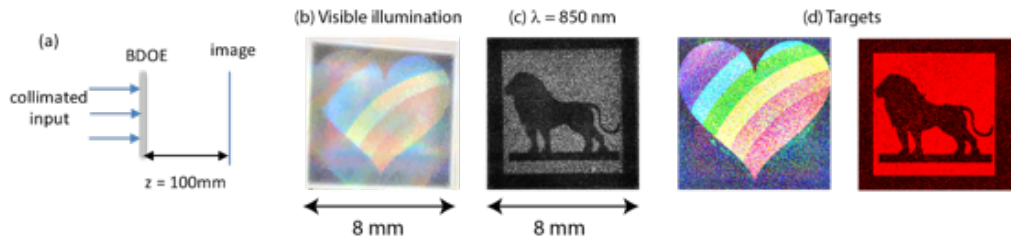

Fig. 2. Different images produced by one BDOE under different wavelengths. (a) Schematic of the experimental setup. Experimental images of (b) a rainbow heart with collimated light at λ = 400 – 700 nm (supercontinuum laser source) and (c) a lion silhouette with collimated light at 850nm (laser source). Target images are shown in (d). Also, see Visualization 1 for observations under sunlight.

In the Fresnel-diffraction regime, wavelength and propagation distance are inter-changeable. Therefore, one would expect the BDOE to be able to project different images in different image planes as illustrated in Fig. 3 under monochromatic illumination [9]. In fact, we previously demonstrated this inter-changeability in experiments with laser illumination [10]. However, this inter-changeability is not obvious for broadband illumination. Here we show that even under broadband illumination, by appropriately designing the BDOE geometry, it is possible to control how the light intensity changes upon free-space propagation. During the design phase, we maximized the imaging efficiency averaged over the multiple (discrete) planes and the multiple spectral bands, simultaneously. When illuminated by collimated super-continuum white light (SuperK EXTREME source and SuperK VARIA filter with $\lambda = 400$nm to 700nm), the BDOE shown in Figs. 3(b-c), projects the image 1 ("+") at a distance of 40mm and the image 2 ("×") at a distance of 60mm (see Visualizations 2 and 5, showing the transition). Another pair of achromatic images (digits "7" and "9") projected by a different BDOE are shown in Fig. 3(e). In this case, the illumination was the white-LED flashlight of a mobile phone placed about 300mm away from the BDOE, rendering the incident light collimated. Some crosstalk between the images in the different planes is observed. The global SSIM (cross-talk) Fig. 3(d) between the two planes was 0.167 and 0.55 in simulation and experiment, respectively [7]. In order to reduce the cross-talk, we can reduce the pixel width and increase the pixel height as that will increase the imaging efficiency. We believe that this can be further mitigated by spacing the multiple planes farther apart.

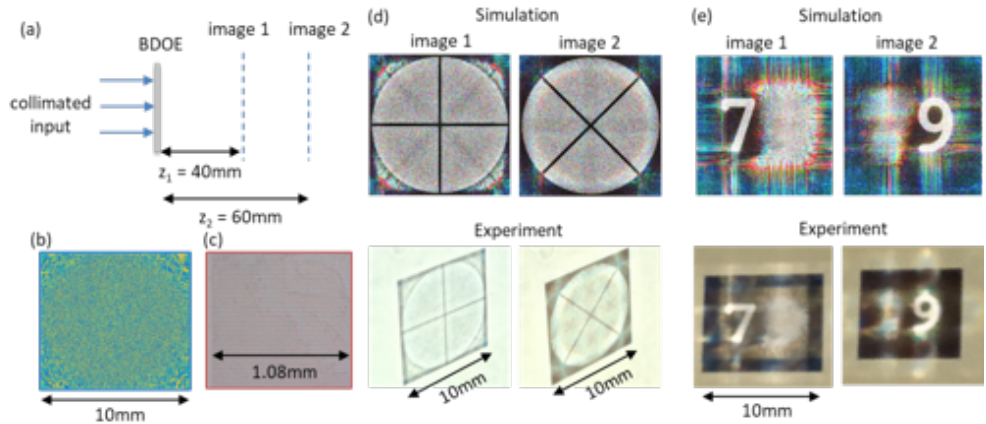

Fig. 3. Two different images produced by one BDOE at two different image planes. (a) Schematic of the experimental setup. (b) Designed height map of the BDOE. (c) Optical micrograph of a small portion of the fabricated BDOE. Simulated (top row) and Experimental (bottom row) images of image 1 and image 2 for the BDOE in (c) are shown in (d) (Visualizations 2 and 5), and for a different BDOE are shown in (e). The illumination was a collimated beam from a super-continuum source ($\lambda = 400$nm to 700nm) and the cellphone flashlight was placed about 300mm away from the BDOE for the "+/x" and the "7/9" devices, respectively. The experimental images were obtained by photographing the image projected onto a white screen.

The approach can be extended to 3 or more planes as well. In Fig. 4, we illustrate the example of 3 different images in 3 different planes (see Visualizations 3 and 6). The illumination was collimated super-continuum white light (SuperK EXTREME source and SuperK VARIA filter ($\lambda=400$nm to 700nm). The global SSIM (crosstalk) Fig. 4(b) between the first two planes was 0.133 and 0.74 in simulation and experiment, respectively, between the last two planes was 0.197 and 0.56 in simulation and experiment, respectively and between the first and last planes was 0.125 and 0.67 in simulation and experiment, respectively [7].

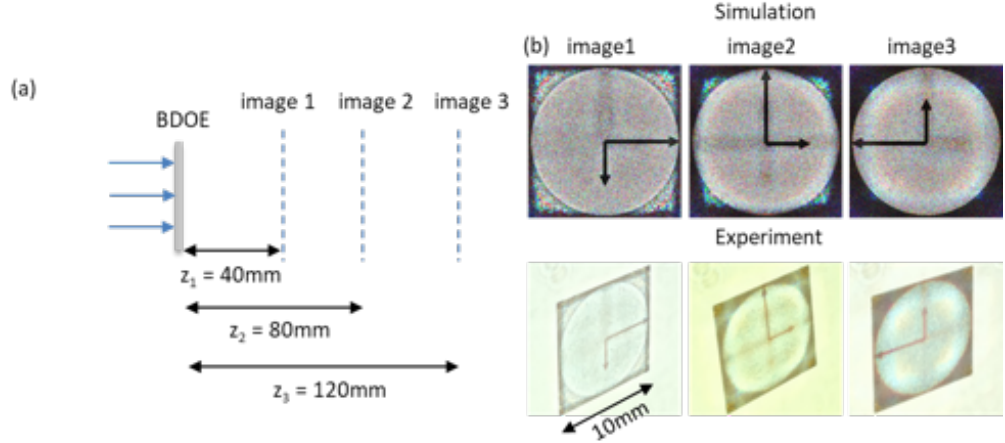

Fig. 4. Three different images produced by one BDOE at three different image planes. (a) Schematic of the experimental setup. (b) Simulated (top row) and Experimental (bottom row) images of image planes 1, 2 and 3. The illumination was a collimated beam from a super-continuum source ($\lambda$ = 400nm to 700nm). The experimental images were obtained by photographing the image projected onto a white screen (Visualizations 3 and 6).

## 3. The BDOE as a Lens

Previously, we have shown broadband diffractive lenses for imaging [11-16]. In analogy with that work, here we show that the BDOE is equivalent to a lens whose point-spread function (PSF) is structured in the form of the desired image. If such a lens is illuminated by collimated source (plane waves), it will form its PSF at a location equivalent to the focal plane. In the case of the BDOE, the desired image is produced at this plane. Elsewhere, the image would be blurred or out-of-focus. Extending this equivalence, when illuminated by a point source (spherical wave), it is expected of such a BDOE to project an image at a distance farther than the focal plane determined approximately by the well-known lens-maker's equation: $1/z_i + 1/z_s = 1/z_o$, where $z_i$, $z_s$ and $z_o$ are the image distance, source distance and "focal" distance all measured from the BDOE (Fig. 5a). In addition, the image formed is a convolution of the magnified PSF and the point source forming the incident spherical wave, where the magnification is determined approximately by $z_i/z_s$. The "focal" distance corresponds to the image distance used during design of the BDOE under collimated illumination. In Fig. 5b, we experimentally show the measured values of $z_s$ for various values of $z_i$ for a BDOE with designed $z_0$ = 38mm and the projected image of a white digit "5." The source comprised of a white LED light with an aperture in front in order to create an approximate point source. Collimation was achieved by placing this source at $z_s$ = 9m, where $z_i \sim z_0$ = 38mm. The images formed by the same BDOE under super-continuum collimated light (SuperK EXTREME and VARIA) and with magnification are shown in Figs. 5(c) and (d), respectively. Therefore, the BDOE is able to project images with magnification without the need for any additional optics. For the images in Fig. 5, we used a 300μm pinhole in front of the LED flashlight to simulate spherical wave illumination. For illumination with super-continuum collimated light, full white spectrum (400nm – 700nm) was used. The illustrations, photographs and further details of the imaging setups provided in ref [7].

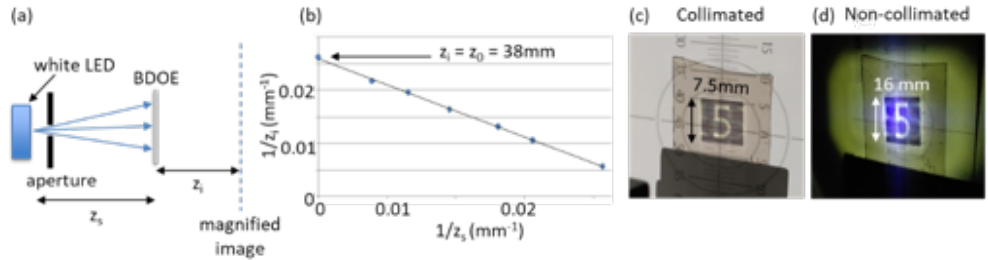

Fig. 5. Lensless image magnification with a BDOE. (a) Schematic of experimental setup. (b) The image formed by the BDOE satisfies the well-known lens-maker's equation illustrating its equivalence to a lens. The image distance under collimated light, $z_0$ is equivalent to the focal length of the lens. Images produced by this BDOE when illuminated by (c) collimated white light and (d) by non-collimated illumination showing magnification of the image (note the change in scale). The source comprised of an opaque screen with a 0.3mm diameter aperture placed in front of a white-LED flashlight (see Visualizations 4 and 7). The distances ($z_s$ and $z_i$) were about (9m and 40mm) and (55mm and 120mm) in Figs. 5(c) and (d), respectively.

## 4. Planar and Reflective BDOEs

BDOEs offer two other advantages. Firstly, one can hide the geometry of the BDOE by utilizing a top-coating as illustrated in Fig. 6(a), creating a flat BDOE. As long as the relative difference in refractive index ($n_1$ vs $n_2$) is non-zero, such devices can be designed in the same manner as conventional BDOEs. A layer of diluted polyvinyl alcohol (Sigma Aldrich) (0.001% by weight in water) was spin coated at 2000 rpm for 60 seconds on top of the fabricated BDOE. We show an example design of a flat BDOE in Fig. 6(b), where the top-coating was polyvinyl alcohol (the patterned region was Shipley 1813 photoresist as in all the previous devices, see ref [7] for additional details.). The simulated and experimental images under collimated white light are shown in Figs. 6(c) and (d), respectively (image distance = 50mm). The image contrast is somewhat lower, but this can be improved by using materials with higher refractive-index contrast or by increasing the maximum pixel depth. It is important to note that since meta-holograms have much higher aspect ratios than BDOEs, applying a top-coating to render them flat is very challenging. Secondly, the minimum feature widths required for BDOEs is limited by diffraction, while meta-holograms require deep sub-wavelength constituent features and concomitantly high aspect ratios that are far more challenging to manufacture.

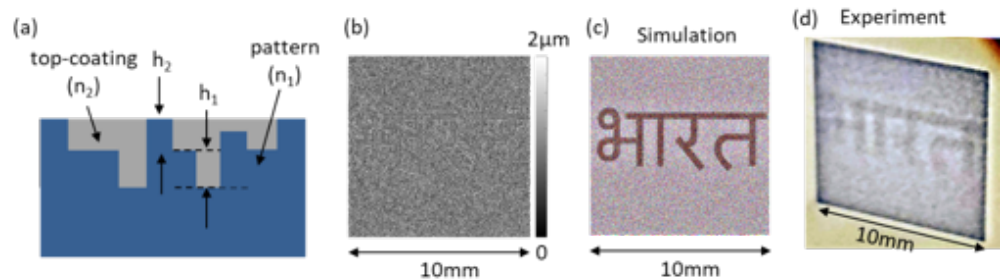

Fig. 6. Flat BDOE. (a) Schematic of a top-coating on patterned layer to create a flat BDOE. (b) Design of a flat BDOE with pixel width of 20μm and maximum pixel depth of 2.6μm. Polyvinyl alcohol was used as the top-coating material. Image of this flat BDOE under white collimated light in (c) simulation and (d) via experiment.

In some applications (for example in packaging), one would require the BDOE to form an image in reflection as shown in the schematic in Fig. 7(a). We designed such a reflective BDOE by first depositing a reflective film (100 nm of Ag was deposited on a pristine glass wafer substrate via sputtering with a 5 nm of Cr layer in between for better adhesion using the Denton

Discovery 18 sputter tool). The parameters of the sputtering process in given in Table S2 [7]. Further details of the sample inspection and related metrology are given in section 3 of ref [7]. Then the photoresist was patterned using grayscale lithography as before (Fig. 7b). The calibration procedure has to be carefully performed to account for the light reflections at the interfaces. Since the light undergoes two passes through the patterned photoresist, the associated phase shift is doubled (at normal incidence). An optical micrograph of a portion of the fabricated device is shown in Fig. 7(c) with pixel width = 20μm. The image formed at a distance of 50mm when illuminated by a mobile-phone flashlight and collimated white laser light, is shown via simulation and via experiment in Figs. 7(d) and 7(e, f) respectively. The experiment was performed at a small incidence angle as illustrated in Fig. 7(a). It is possible to obtain full color images in this scenario if one takes into account the reflection spectrum of the reflective film used. In our preliminary demonstration, we did not account for this effect. Nevertheless, we showed good achromatic image projection by capturing the images at narrowband and broadband illumination using a tunable super-continuum source (see Figs. 7f-i and in ref [7]). As far as we are aware, this is the first demonstration of a reflective image-projecting hologram that is achromatic over the visible band (450nm to 700nm).

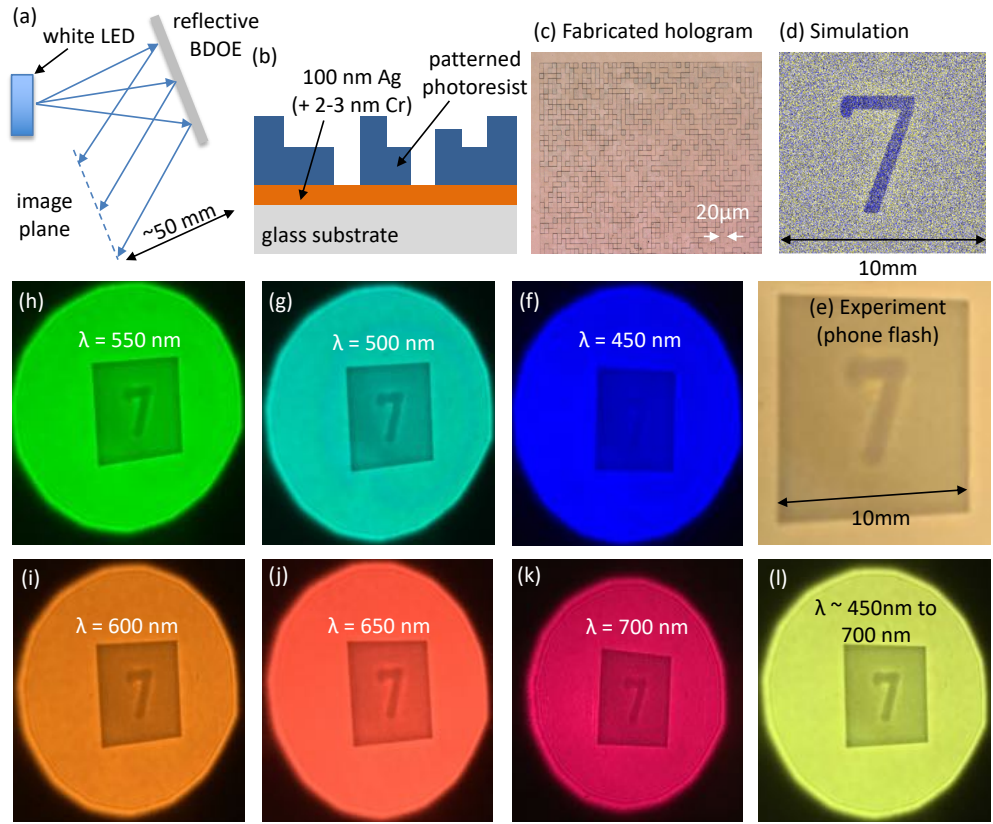

Fig. 7. Reflective BDOE. (a) Schematic showing the BDOE forming an image in reflection. (b) Schematic of the reflective BDOE microstructure. (c) Magnified optical micrograph of a portion of the fabricated reflective BDOE showing good pixel fidelity. (d) Simulated and (e - l) experimental image formed when illuminated by (e) a mobile-phone flashlight and (f-l) collimated laser light at normal incidence (close to normal incidence for experiment) for different wavelengths (bandwidth = 50 nm). The illumination sources were (e) a white-LED flashlight from a mobile-phone placed about 300mm away from the BDOE and (f-l) collimated tunable super-continuum source.

In Fig. 8, we showcase the color multiplexing capability of a transmission BDOE that projects the image of a rainbow heart (when illuminated with visible light) and lion-silhouette (when illuminated with NIR light). It can be seen that the same device projects the rainbow heart pattern in the visible band and the lion silhouette in the NIR band. It is interesting to note that the lion silhouette starts appearing at around 700 nm illumination and becomes dominant in till 900 nm (Images up to 950 nm illumination are shown in ref [7]).

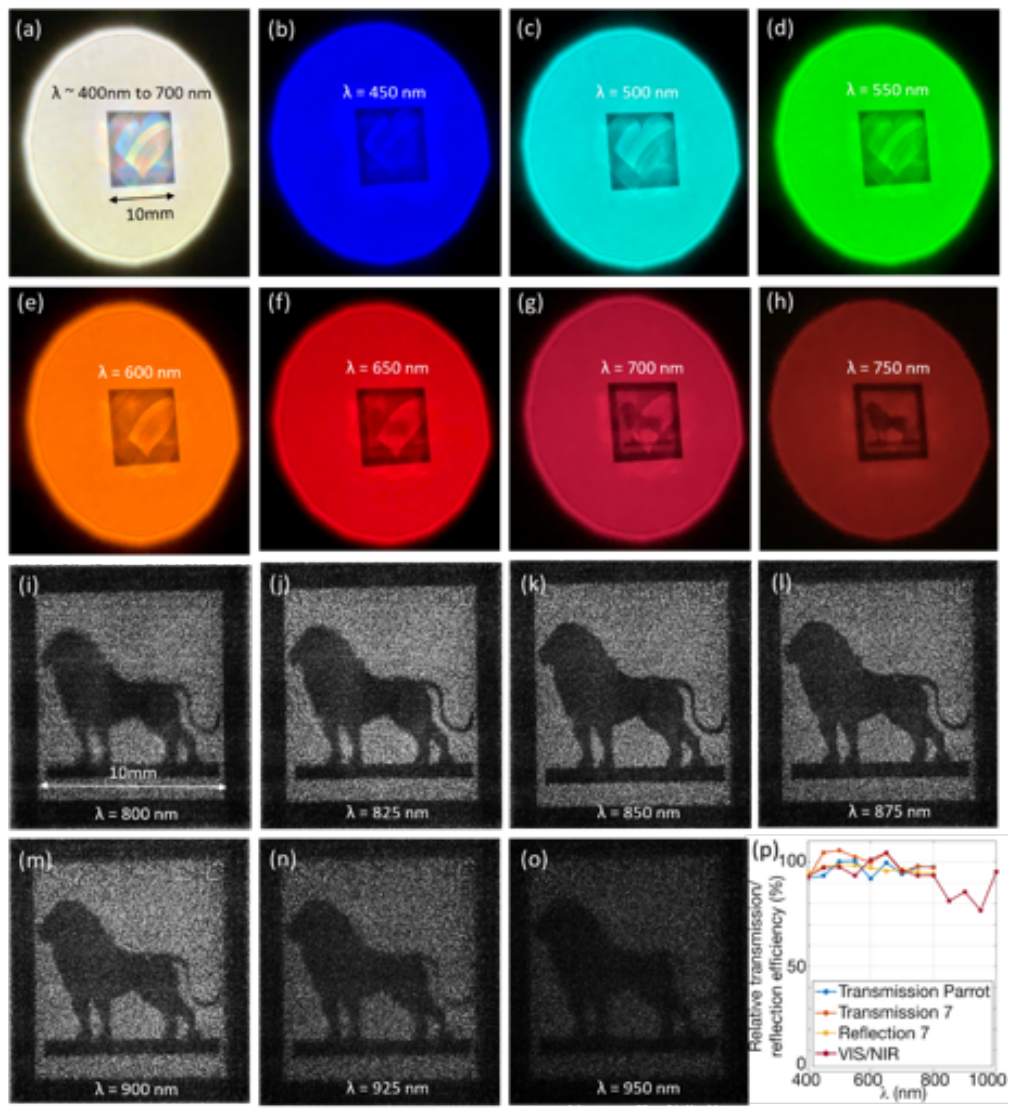

Fig. 8. Color-multiplexed images in a transmission rainbow heart (visible) and lion-silhouette (NIR) BDOE and Transmission/Reflection efficiency measurements. (a-o) Images produced by the BDOE when illuminated by collimated tunable supercontinuum source with (a) broadband (400-700 nm) and (b-o) narrowband (bandwidth = 50nm) illumination. The images (i-o) were captured by projecting the image on a CMOS sensor since the camera used to capture images (a-h) was not capable of photographing in the IR band. In (a), minimal tone adjustment was performed to remove an extraneous yellow tint from the image that was present in the photograph, but not seen with the naked eye. No other image processing has been performed. (p) Relative transmission/reflection efficiency spectra of the BDOE (VIS/NIR) shown in (a-o), reflective BDOE shown in Fig. 7 (Reflection 7), and other transmissive BDOE (Transmission parrot and 7, not shown here) that project images of a parrot and a "7" in transmission. Details of the measurements are in ref [7].

Excellent color fidelity is observed under both the broadband (a) and narrowband illumination (b-o). We also measured the transmission efficiency of 4 different BDOEs and the results are summarized in Fig. 8(p). The relative transmission/reflection efficiency is measured as the ratio of power transmitted/reflected by the BDOE to the power transmitted/reflected by a reference (unpatterned) device. As was pointed out previously [4], the multilevel microstructures function as an anti-reflection coating, which enables the relative efficiency to go above 100% at certain wavelengths. The average transmission efficiencies from 400nm to 800nm is over 96% for all 4 measured BDOEs and ~ 85% for the VIS/NIR BDOE in the 850-1000 nm range. Details of these measurements are included in ref [7]. We emphasize that transmission efficiency is somewhat unrelated to the imaging efficiency. The latter measures the fidelity of the projected images, while the former simply states how much power is transmitted.

## 5. Conclusions

In summary, here we demonstrate that appropriate design of multi-level diffractive structures can enable efficient image projection over broadband and multiple spectral bands, and onto multiple image planes. Since such structures may be fabricated at low cost over large areas and at high speed via various imprinting technologies, [17] these can be readily used in anti-counterfeiting applications. We note that nano- and micro-optical elements are already prevalent in huge volumes in many consumer electronics and banknotes [18]. In addition, we point out that metamaterials and metasurfaces for image projection does not offer any advantage over BDOEs. In fact, BDOEs are far easier to manufacture due to (1) their much larger feature widths, (2) much smaller aspect ratios and (3) their use of low-index polymers that are readily amenable to imprinting. Since spatial frequencies that are larger than $2\pi/\lambda$ do not propagate in free space, there is no need for pixel widths smaller than $\lambda/2$. Their differences have been discussed in some details in ref [11] in the context of imaging. This also means that even though meta-holograms have pixels with widths far smaller than the wavelength, the free-space propagation cutoff limits information transfer to the same spatial frequencies as in the case of BDOEs. In other words, meta-holograms cannot transmit any additional information than BDOEs into the regions of space farther than a few wavelengths away. We do note that meta-holograms do have unique advantages over BDOEs in the case of polarization- or spin-encoded holography [19–22]. But these suffer from poor efficiency and require sub-100nm structures, which is challenging to manufacture over large areas.

### Acknowledgements

We thank Brian Baker, Steve Pritchett and Christian Bach for fabrication advice, and Christian Skipper (Keyence) for providing the 3D optical micrograph.

### Funding

Office of Naval Research grant N66001-10-1-4065. Utah Science Technology and Research (USTAR) Initiative research grant.

### Author Contributions

R.M. and M.M. conceived and designed the experiments, modeled and optimized the devices. M.M. fabricated the devices. M.M and A.M. performed the experiments and numerical analysis. All authors analyzed the data and wrote the paper.

### Competing Interests

R.M. is co-founder of Oblate Optics, which is commercializing the subject technology. The University of Utah has applied for a patent covering the subject technology.

# Supplementary Information

# Multi-level diffractive-optics enables versatile, multi-plane holography with temporally-incoherent illumination

Monjurul Meem[1], Apratim Majumder[1], and Rajesh Menon[1,2,*]

[1]Department of Electrical and Computer Engineering, University of Utah, Salt Lake City UT 84112

[2]Oblate Optics, Inc., 13060 Brixton Place, San Diego CA 92130

*rmenon@eng.utah.edu

## 1. Design and geometric parameters of the holograms

The holograms are pixelated in X and Y directions, height of each pixel is quantized into multiple levels. Different parameters have been used for the hologram designs. These are summarized in Table S1.

*Table S1. Design and geometric parameters of fabricated holograms.*

| Hologram demonstrated in | Pixel size (µm) | Number of pixels | Physical size (mm X mm) | Maximum height (µm) | Propagation distance (mm) | Operating wavelengths (nm) | η |
|---|---|---|---|---|---|---|---|
| Figs. 1, 2, 8 | 20 | 500 x 500 | 10 x 10 | 2.6 | 100 | 400-700, 850 | 64% |
| Fig. 3 | 20 | 500 x 500 | 10 x 10 | 2.6 | 40, 60 | 400 - 700 | 29% |
| Fig. 4 | 20 | 500 x 500 | 10 x 10 | 2.6 | 40, 80, 120 | 400 - 700 | 91% |
| Fig. 5 | 10 | 750 x 750 | 7.5 x 7.5 | 2.6 | 38 | 400 - 700 | 48% |
| Fig. 6 | 20 | 500 x 500 | 10 x 10 | 2 | 100 | 400 - 700 | 95% |
| Fig. 7 | 20 | 500 x 500 | 10 x 10 | 1.3 | 50 | 400 - 700 | 60%* |

* not including reflectivity of metal.

The wavelength-averaged imaging efficiency was calculated using the equation below as has been described previously:

$$\eta = \frac{1}{N}\Sigma_\lambda \frac{\Sigma_m \Sigma_n I_T^{(\lambda)} |U(p_{m,n})|^2}{P_{in}^{(\lambda)}},$$

where, $\eta$ is the wavelength averaged imaging efficiency, $\lambda$ is the design wavelength, $N$ is the number of design wavelengths, $I_T^{(\lambda)}$ is the target image pattern at wavelength $\lambda$, $U(p_{m,n})$ is the complex amplitude at the reconstruction plane diffracted by the hologram with height profile distribution $p_{m,n}$, $m$ and $n$ are the pixel indisces, and $P_{in}^{(\lambda)}$ is the input power at wavelength $\lambda$. The objective of the optimization is to determine a height profile ($p_{m,n}$) so that the wavelength averaged diffraction efficiency is maximized. For the multi-plane devices, the efficiency was further averaged over the different image planes.

## 2. Fabrication

The holograms were fabricated using direct laser write grayscale lithography [1]. A positive tone photoresist (s1813) [2] was spin coated on a 2" soda lime glass wafer at 1000 rpm for 60 seconds to yield a thickness a 2.6 μm. The spin coated sample was baked in an oven at 110°C for 30 minutes. The designs were written on the sample using "Heidelberg μPG 101" tool [3] and developed in AZ 1:1 solution for 35 seconds. A calibration step was performed beforehand to determine the exposed depths at a particular gray scale level. The details of the fabrication process have been discussed elsewhere [4].

For the hologram shown in Fig. 7, a diluted polyvinyl alcohol (0.001% water) was spin coated at 2000 rpm for 60 sec. For the reflection hologram (Fig. 8), a 100 nm of Ag was deposited on the sample via sputtering with a 5 nm of Cr layer in between for better adhesion, the rest is same as aforementioned process. The parameters of the sputtering process in given in table S2

*Table S2. Process parameters of Ag deposition via sputtering.*

| Tool | Denton Discovery 18 |
| --- | --- |
| substrate | Glass |
| base-pressure (10^-6 torr) | 2 |
| power-supply | DC |
| sputter-material-1 | Cr |
| pre-sputter-time-1 (minutes) | 2 |
| sputter-time-1 (minutes) | 0.3333333 |
| sputter-power-1 (W) | 100 |
| argon-pressure-1 (mT) | 4.4 |
| argon-flow-1 (%) | 50 |
| o2-flow-1 (%) | 0 |
| sputter-material-2 | Ag |
| pre-sputter-time-2 (minutes) | 2 |
| sputter-time-2 (minutes) | 1.5 |
| sputter-power-2 (W) | 100 |
| argon-pressure-2 (mT) | 4.4 |
| argon-flow-2 (%) | 50 |
| o2-flow-1 (%) | 0 |

## 3. Effect of fabrication error

The fabricated holograms do not exactly match with the design heights due to non-repeatability in grayscale process. The pixel heights of the fabricated holograms were measured using a stylus profilometer (Tencor P-10). Only the top left row was measured for simplicity. The results along with the design values of the hologram demonstrated in Fig. 1, Fig. 2 and Fig. 8 (all three represent the same BDOE) is illustrated in Fig. S1 as an example. The estimated an average and standard deviation of error in pixel heights are summarized in table S3.

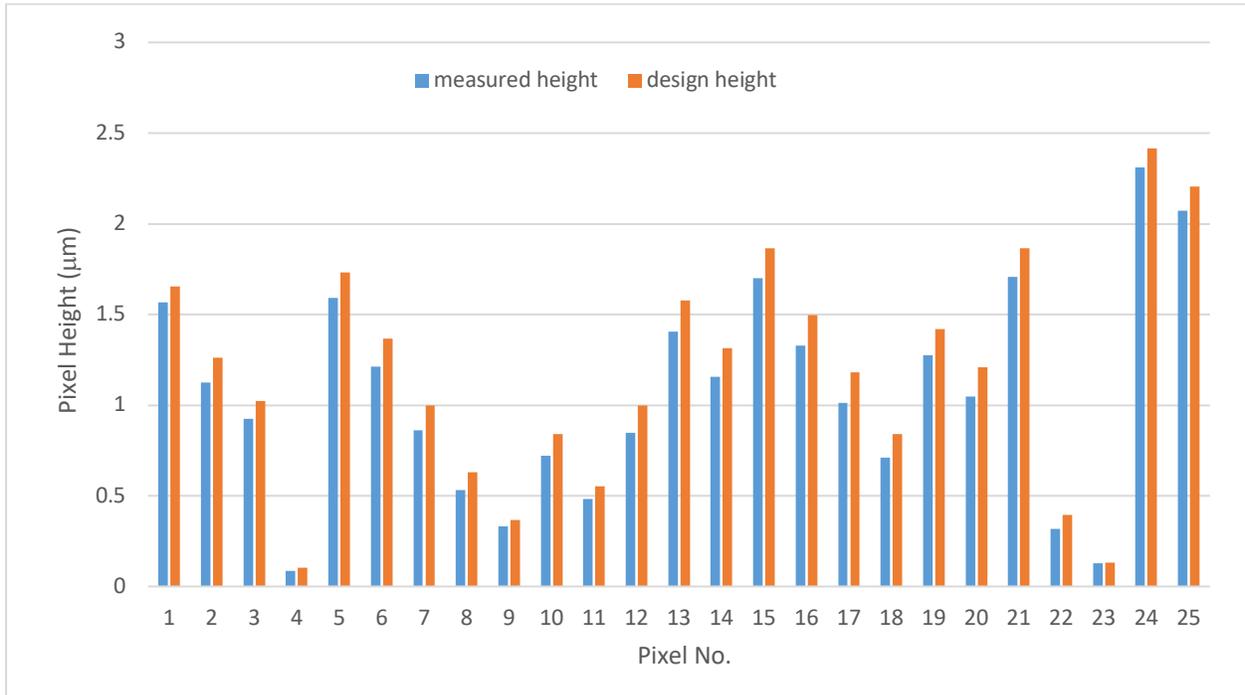

*Fig. S1: Measured pixel heights compared to design values of pixels for hologram demonstrated in Fig. 1, Fig. 2 and Fig. 8*

*Table S3. Pixel height errors of the fabricated holograms.*

| Hologram demonstrated in | Average error (nm) | Standard deviation of error (nm) |
|---|---|---|
| Fig. 1, 2 & 8 | 60 | 48 |
| Fig. 3 (d) | 68 | 65 |
| Fig. 3 (e) | 85 | 58 |
| Fig. 4 | 38 | 33 |

To investigate the effect of fabrication error, random heights drawn from a normal distribution with zero mean and different standard deviations were added to the design heights and corresponding imaging efficiency were simulated. For simplicity, only the result for BDOE in Fig. 1, Fig. 2 and Fig. 8 (same BDOE) is presented. As expected, the efficiency decreases as the standard deviation of the height errors increase (Fig S2). The resulting effect on the reconstructed images is shown in Fig. S3. The reconstructed image shows good fidelity up to 100nm of standard deviation of error.

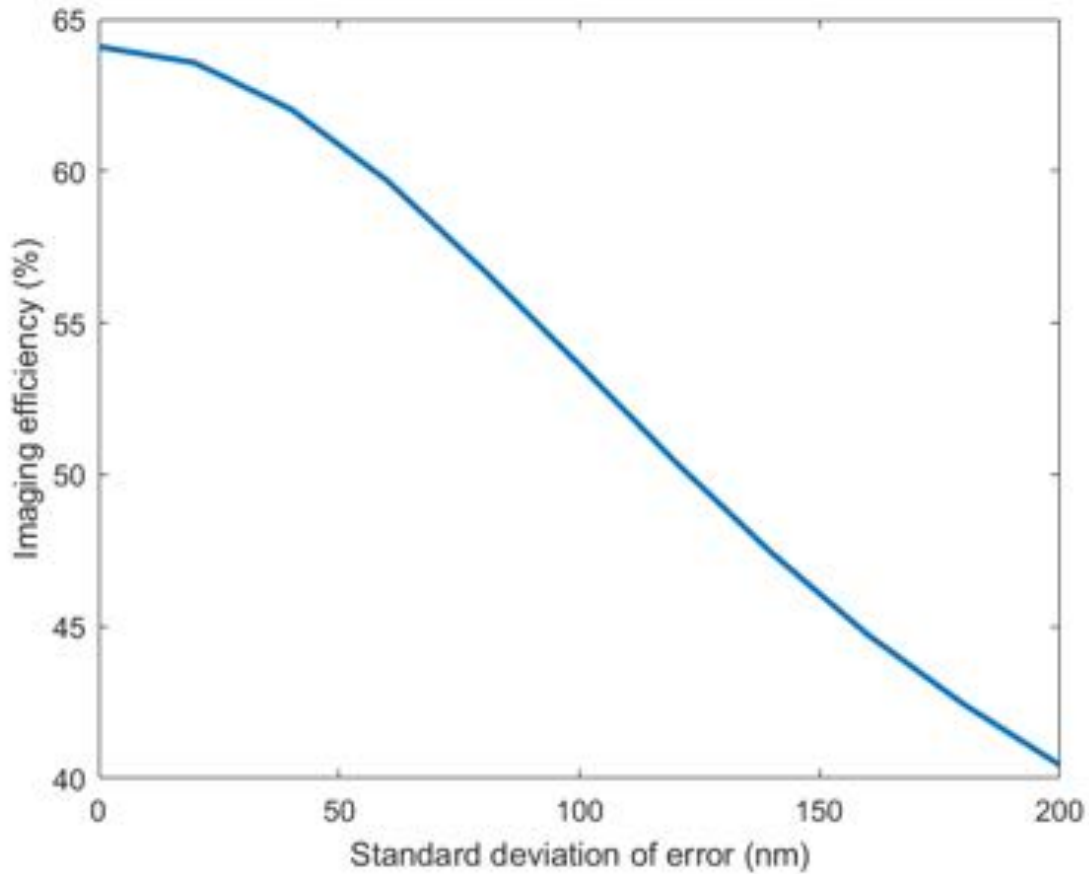

*Fig. S2: Effect of fabrication error on imaging efficiency: Efficiency as a function of random height errors with zero mean and different standard deviations.*

**3. Imaging setup**

For characterizing the hologram, the image was projected/reflected onto a white screen and the reflected image was photographed. We used two different illumination sources. One was collimated and expanded beam from SuperK VARIA filter [6] which in turns was connected to SuperK EXTREME EXW-6 source [7] while the other was a white LED from flashlight or a mobile phone flashlight. For the images in Fig. 5, we used a 300μm pinhole in front of the LED flashlight to simulate spherical wave illumination. For illumination with SuperK VARIA, full white spectrum (400nm – 700nm) was used. The imaging setups are illustrated in Fig. S4.

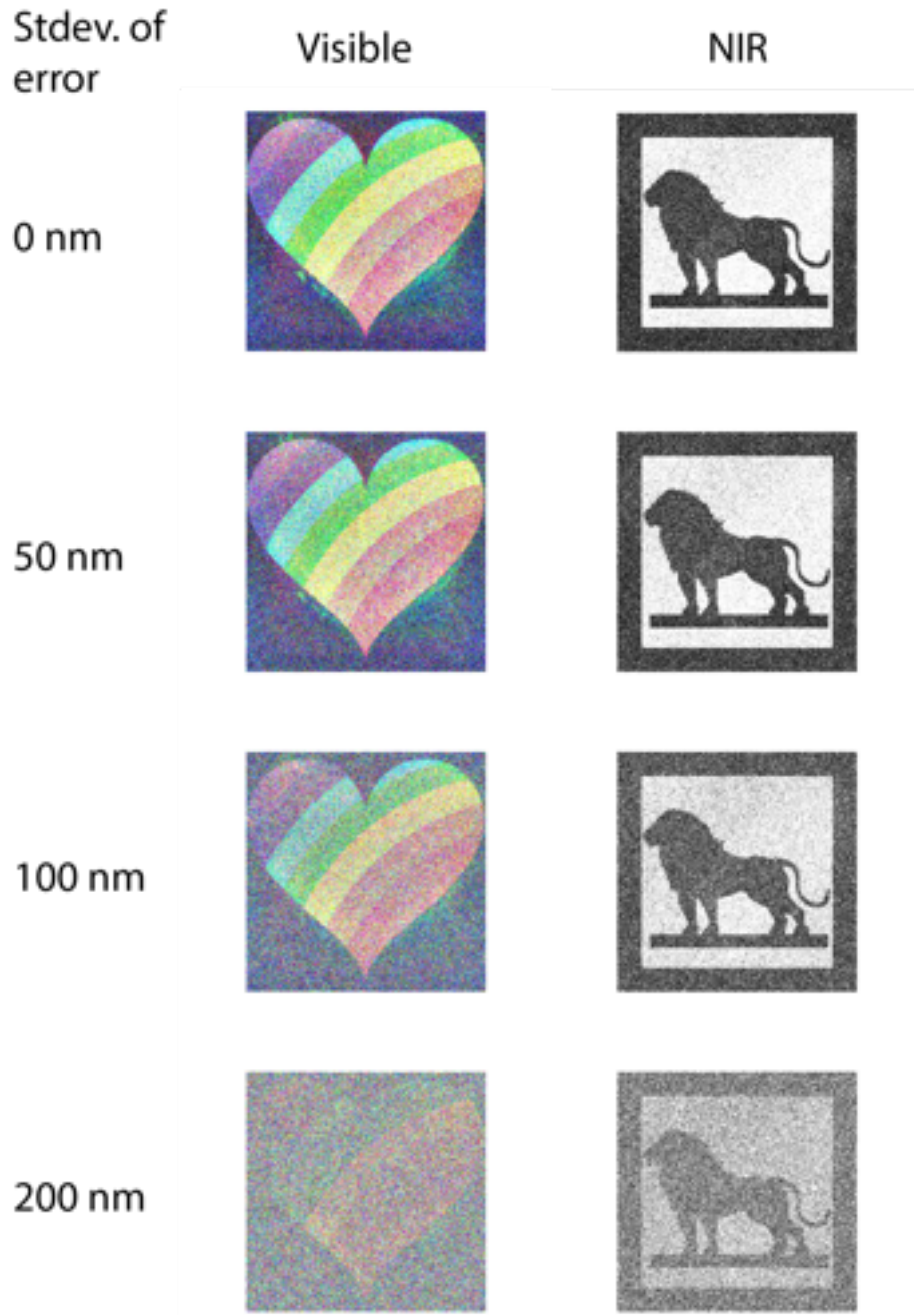

Fig. S3: Reconstructed images at different standard deviations of height errors.

## 5. Relative transmission/reflection efficiency measurement setup

The experimental setup shown in Fig. S5 was used to measure the relative transmission efficiency of the holograms. In transmission, we placed an aperture of the same size as the hologram in front of the hologram (see Fig. S5) as well as a substrate with photoresist that is unpatterned, to serve as reference. The relative transmission efficiency was defined as the ratio of the power transmitted when the hologram is placed with respect to that when the unpatterned substrate is placed. In reflection mode, everything was the same except that the aperture was moved before the device, thereby facing the incident light.

**6. Imaging efficiency**

The imaging efficiency of the BDOE shown in Fig. 1, Fig. 2 and Fig. 8 (same BDOE) were calculated form the raw images captured on an image sensor (DFM 72BUC02-ML, The Imaging Source). The schematic of the experimental setup is shown in Fig. S6. The BDOE was illuminated with collimated beam at four different wavelengths (450nm, 550nm, 650nm and 850nm) with 15nm bandwidth. For each illumination wavelength, corresponding reconstructed image was captured by the color sensor. In each case, a dark image was also recorded and subtracted from the reconstructed images.

**7. Supporting Media.**

Supplementary Video 1 shows the image projected under sunlight using the device shown in Fig. 2. It shows that by placing a visible-cut filter, the IR image of the lion becomes visible.

Supplementary Video 2 shows the projected image changing as the screen is moved away from the BDOE described in Fig. 3.

Supplementary Video 3 shows the projected image changing as the screen is moved away from the BDOE described in Fig. 4.

Supplementary Video 4 shows the projected image changing magnification as the point source is moved away from the BDOE described in Fig. 5. In this case, the BDOE -screen distance is not changed.

Supplementary Video 5 shows the projected image changing as the screen is moved away from the BDOE described in Fig. 3, taken in better ambient conditions.

Supplementary Video 6 shows the projected image changing as the screen is moved away from the BDOE described in Fig. 4, taken in better ambient conditions.

Supplementary Video 7 shows the projected image changing as the screen is moved away from the BDOE described in Fig. 5, taken in better ambient conditions.

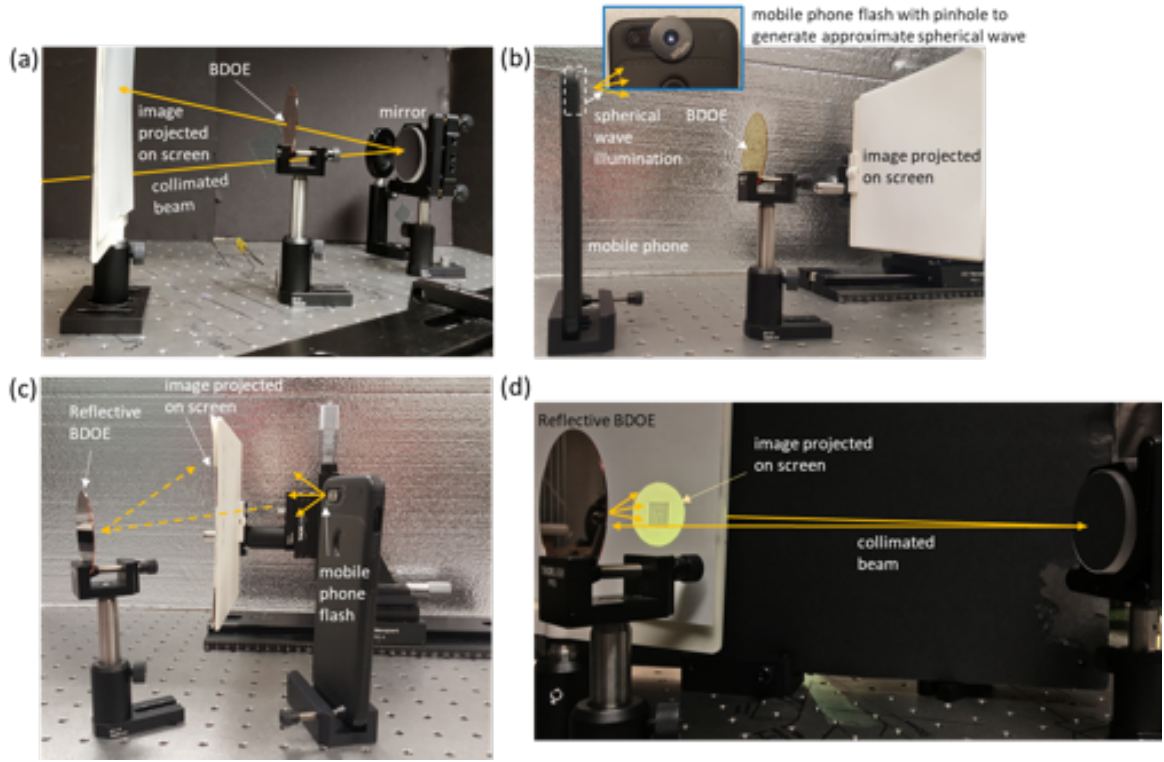

*Fig. S4. Imaging setups for (a) collimated light illumination using white laser light from supercontinuum Super-K Extreme+Varia, (b) mobile phone flash illumination (The inset in (b) shows the pinhole in front of the mobile phone flash for simulating spherical wave illumination and (c) reflective hologram illuminated form front and its image projected onto a screen and (d) reflective hologram illuminated form front using collimated laser light illumination and its image projected onto a screen for generating images shown in Fig. 8.*

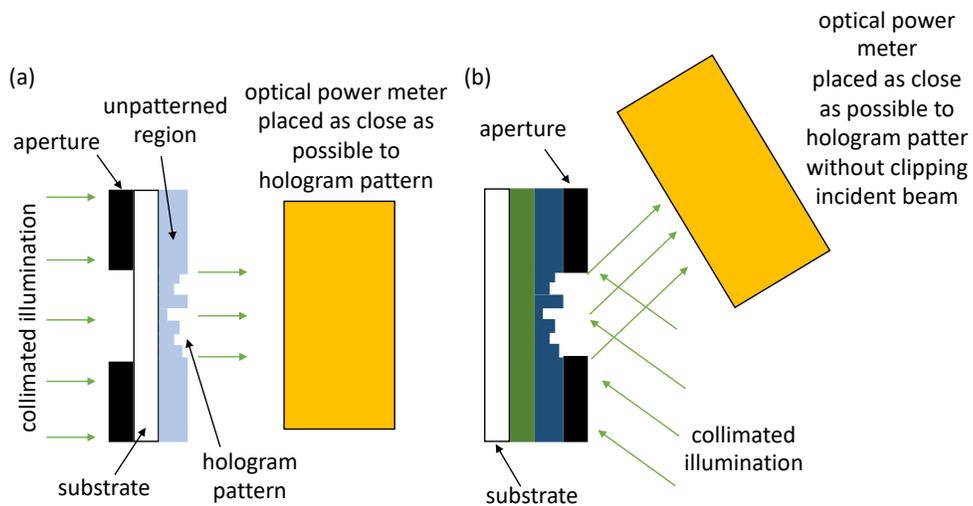

*Fig. S5. Experimental setup for measuring relative transmission efficiency of hologram in transmission mode. For measuring the reference, the hologram was replaced by a substrate with a completely unpatterned photoresist layer.*

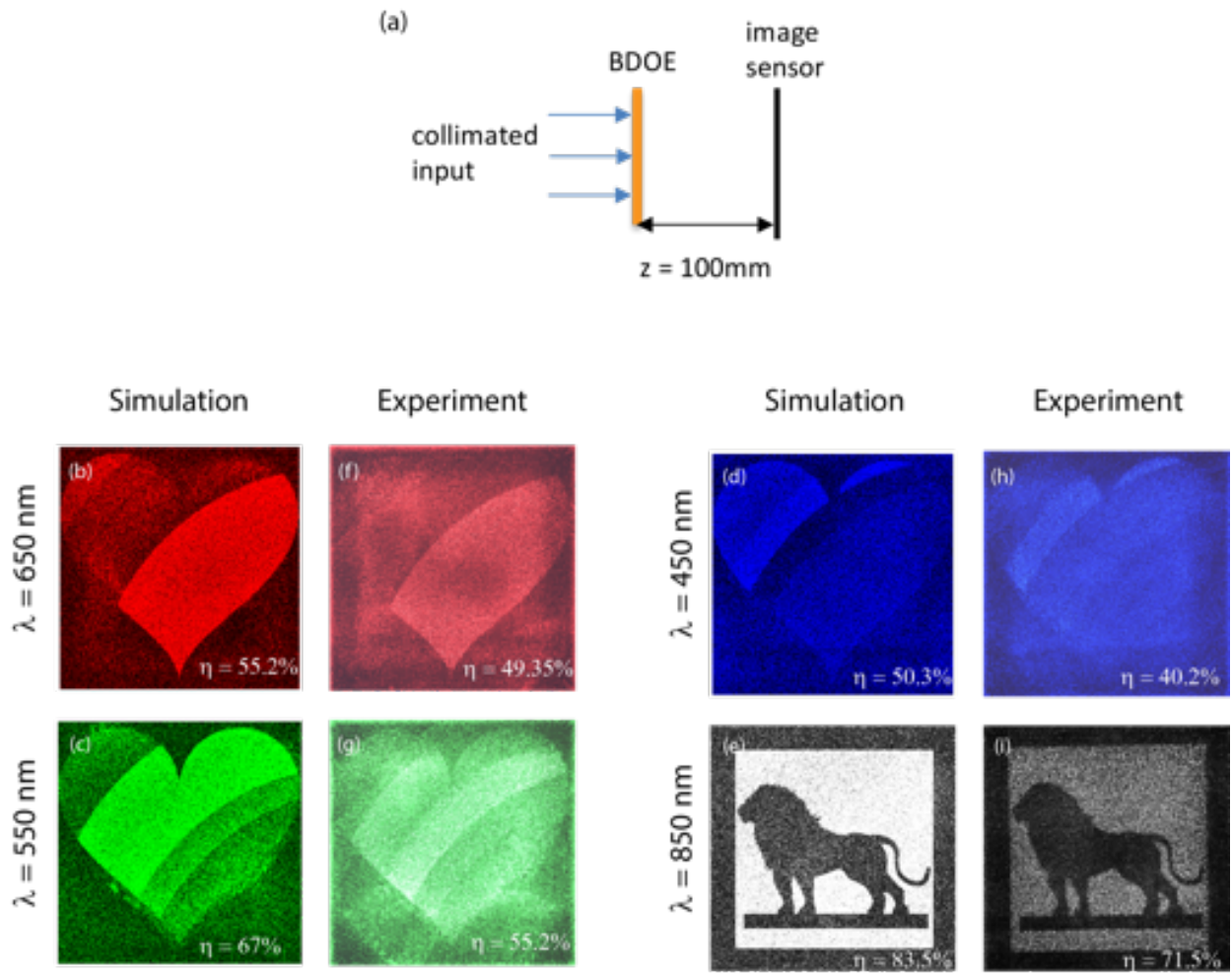

Fig. S6. Imaging efficiency of the BDOE, presented in Fig. 1, Fig. 2 and Fig. 8 (same BDOE) (a) schematic of the experimental setup. The BDOE was illuminated with collimated laser light from a supercontinuum source (NKT Photonics SuperK Extreme + Varia) at different wavelengths and the projected image was captured on an image sensor (DFM 72BUC02-ML, The Imaging Source). Simulated images (b-e) and captured images (f-i) are shown. The simulated and measured imaging efficiency are noted in each image.

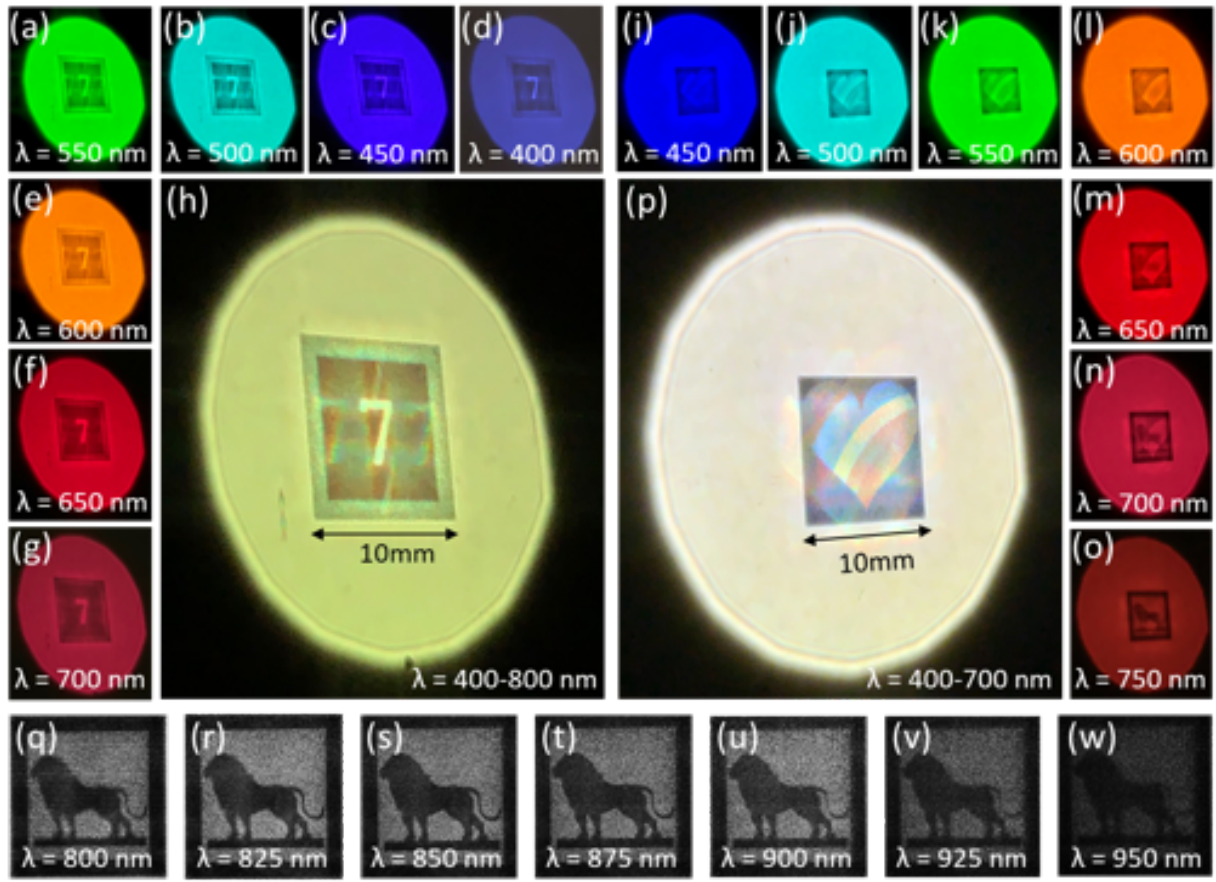

*Fig. S7. Imaging demonstrations of two transmission holograms using collimated white laser light from a supercontinuum source (NKT Photonics SuperK Extreme + Varia) at different narrowbands (bandwidth = 50 nm) (a-g, i-o and q-w) and under broadband illumination (h and p). Images (i-o), (p) (q-w) are all produced by the same BDOE device. The rainbow heart appears in the visible band (400-700 nm) while the lion silhouette appears in the NIR band (750-950 nm).*

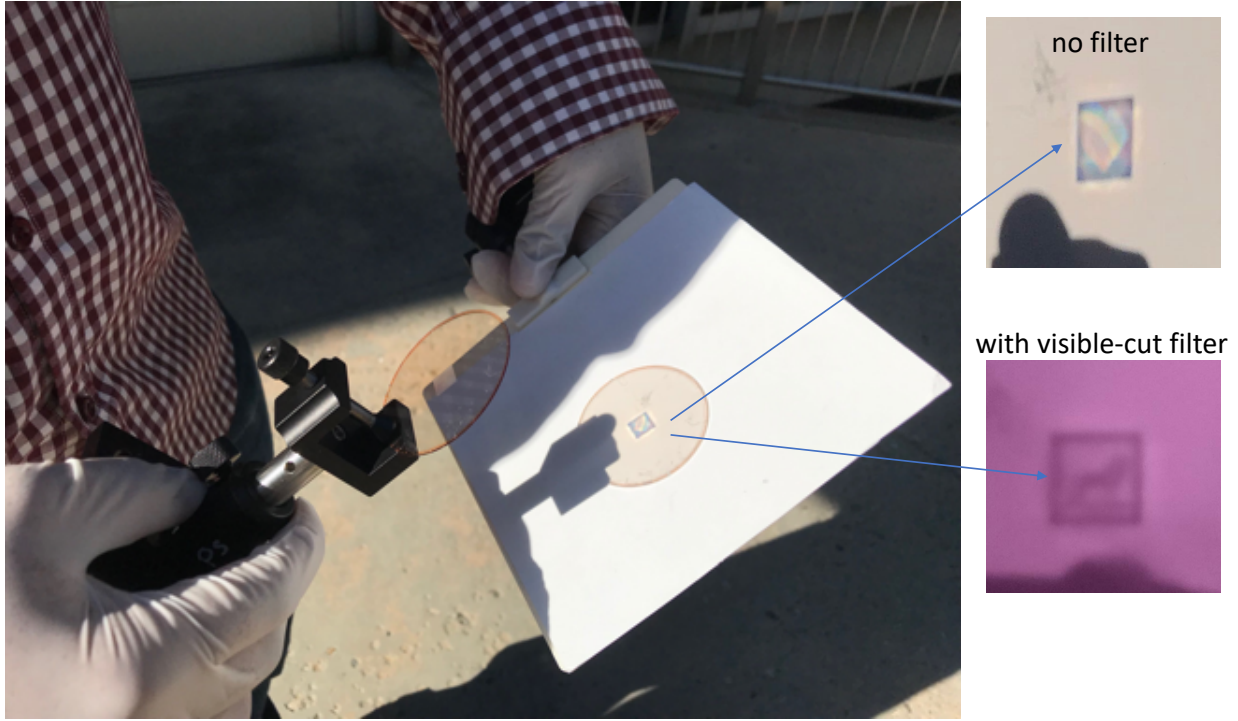

*Fig. S8. Image projection under direct sunlight. These photographs were taken around 6pm on June 18, 2019 in Salt Lake City, UT. The colorful image of the heart is clearly visible. By placing a visible-cut filter, the lion image becomes visible as well.*

**8. Analyzing crosstalk.**

In order to quantify the cross-talk between the different images or the different planes of the images produced by the BDOEs, we compute the Structural Similarity Index (SSIM) [8] which is a popular method of comparing the similarity between two images, which for our case, translates to cross-talk. In Fig. S9, we calculate the local SSIM map for two planes and also the global SSIM value for the case. This is done for both simulations and experiments for the multi-band and multi-plane image projecting BDOEs described in the manuscript.

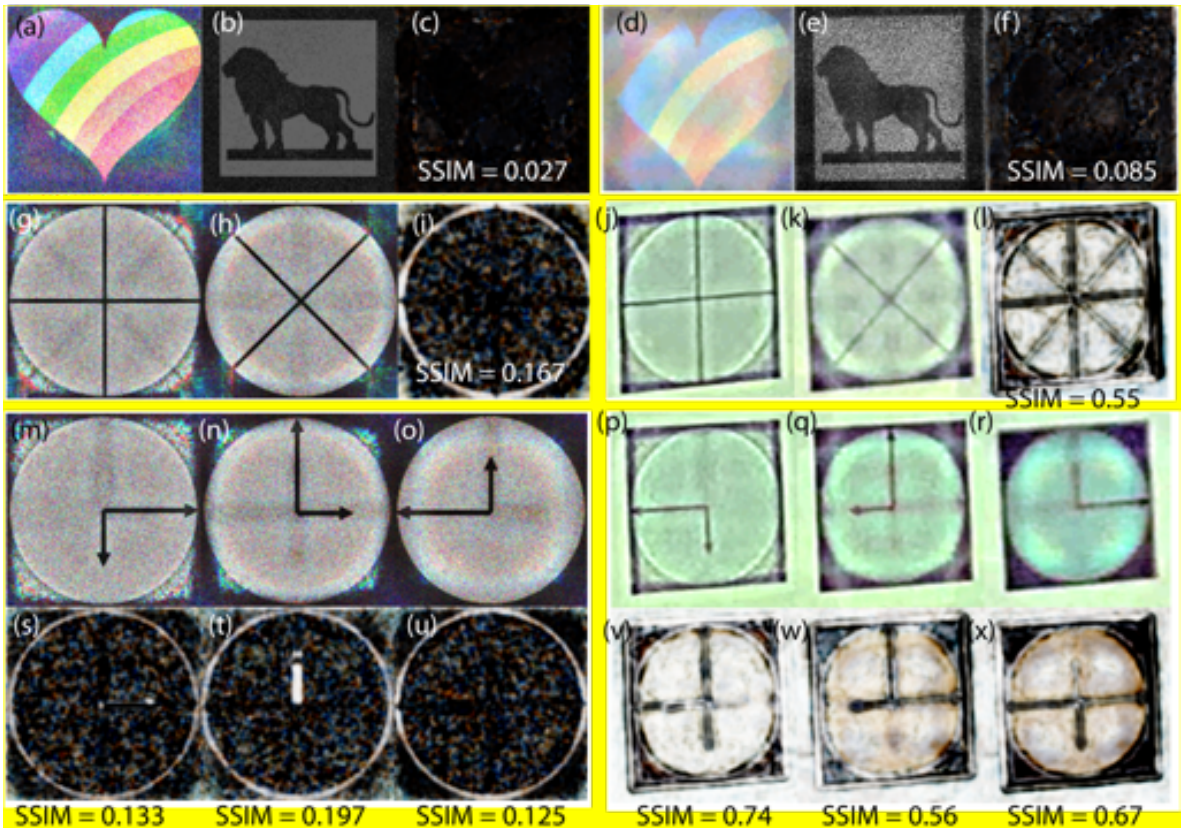

Fig. S9. Crosstalk analysis: (a) Simulated rainbow heart image produced in visible light and (b) lion image produced in NIR by the BDOE device shown in Fig. 2 and (c) the corresponding local SSIM map with the global SSIM value for (a) and (b) equal to 0.027. (d) Experimentally projected rainbow heart image produced in visible light and (e) lion image produced in NIR by the BDOE device shown in Fig. 2 and (f) the corresponding local SSIM map with the global SSIM value for (d) and (e) equal to 0.085. (g, h) Simulated images produced in different planes by the BDOE device shown in Fig. 3(d) and (i) the corresponding local SSIM map with the global SSIM value for (g) and (h) equal to 0.167. (j, k) Experimentally projected images produced in different planes by the BDOE device shown in Fig. 3(d) and (l) the corresponding local SSIM map with the global SSIM value for (j) and (k) equal to 0.55. (m-o) Simulated images produced in different planes by the BDOE device shown in Fig. 4 and (s-u) the corresponding local SSIM maps with the global SSIM values for planes in (m) and (n), (n) and (o) and (m) and (o), equal to 0.133, 0.197 and 0.125 respectively. (p-r) Experimentally projected images produced in different planes by the BDOE device shown in Fig. 4 and (v-x) the corresponding local SSIM maps with the global SSIM values for planes in (p) and (q), (q) and (r) and (p) and (r), equal to 0.74, 0.56 and 0.67 respectively.